# Low bend loss waveguides enable compact, efficient 3D photonic chips

Alexander Arriola,[1,2,3*] Simon Gross,[1,2] Nemanja Jovanovic,[1,4,5] Ned Charles,[6] Peter G. Tuthill,[6] Santiago M. Olaizola,[3] Alexander Fuerbach,[1,2] and Michael J. Withford[1,2,4]

[1]*MQ Photonics Research Centre, Dept. of Physics and Astronomy, Macquarie University, NSW 2109, Australia*
[2]*Centre for Ultrahigh-bandwidth Devices for Optical Systems (CUDOS), Australia*
[3]*CEIT and Tecnun, University of Navarra, Pº Manuel de Lardizabal 15, Donostia-San Sebastian, 20018, Spain*
[4]*Macquarie University Research Centre in Astronomy, Astrophysics and Astrophotonics, Dept. of Physics and Astronomy, Macquarie University, NSW 2109, Australia*
[5]*Australian Astronomical Observatory (AAO), PO Box 296, Epping NSW 1710, Australia*
[6]*Sydney Institute for Astronomy (SIFA), School of Physics, University of Sydney, 2006, Australia*
[*]*alex.arriola@mq.edu.au*

**Abstract:** We present a novel method to fabricate low bend loss femtosecond-laser written waveguides that exploits the differential thermal stabilities of laser induced refractive index modifications. The technique consists of a two-step process; the first involves fabricating large multimode waveguides, while the second step consists of a thermal post-annealing process, which erases the outer ring of the refractive index profile, enabling single mode operation in the C-band. By using this procedure we report waveguides with sharp bends (down to 16.6 mm radius) and high (80%) normalized throughputs. This procedure was used to fabricate an efficient 3D, photonic device known as a "pupil-remapper" with negligible bend losses for the first time. The process will also allow for complex chips, based on 10's - 100's of waveguides to be realized in a compact foot print with short fabrication times.

## References and links

## 1. Introduction

In 1996, *Davis et al.* [1] first demonstrated that a tightly focused femtosecond laser pulse could induce a highly localized and permanent refractive index change in a transparent material, which could then be used to fabricate waveguides. Since then the field of femtosecond laser direct-writing [2-4] has received growing attention from a variety of real-world applications including optofluidics [5-7], quantum optics [8,9] and astrophotonics [10-11]. The femtosecond laser direct-write process can be divided into two different regimes. Using low pulse repetition rates (kHz), a repetitive modification of the material occurs typically creating a Gaussian-like refractive index profile in commonly used materials such as fused silica [12]. This refractive index profile is ideal for guiding light as its behavior is similar to a step-index profile, but the drawback is that for complex devices long fabrication times are required (1-100 hours) due to the low sample translation speeds (typically tens of microns per second) involved. In contrast, by using high pulse repetition rates (MHz) an accumulation of heat occurs because the inter-pulse spacing is shorter than the thermal diffusion time of the material, resulting in local melting, strong heat diffusion followed by rapid quenching of the material due to the high translation speeds (typically millimeters per second). Under these conditions, fabrication times are greatly reduced. However the high repetition rate regime typically creates complex refractive index profiles in multicomponent silicate glasses (which are most commonly used in this regime [13-14]). Figure 1 shows a micrograph image of a typical single-mode waveguide at 1550 nm written with high repetition rates in alkaline earth boro-aluminosilicate glass (Corning Eagle2000). The mode field diameter (MFD) is 10.7 × 10.0 μm which is a close match to the nominal 10.4 μm MFD of SMF-28 fiber.

Typically, such index profiles exhibit a Gaussian-like central region of modification with a strong positive index contrast (bright white region in Fig. 1(a)) surrounded by a uniform ring with a lower but still positive index contrast [13-15] and a depressed cladding region (i.e. a region with a negative index change) in between (dark region in Fig. 1(a)). Such index profiles work well for straight waveguides but exhibit large losses for tightly bent waveguides compared to a step-index waveguide with similar MFD [16]. The requirement of large bend radii in order to obtain a high throughput imposes a limit on the complexity and the footprint of the circuit that can be realized. This fundamental shortcoming motivated the research leading to the solution presented here.

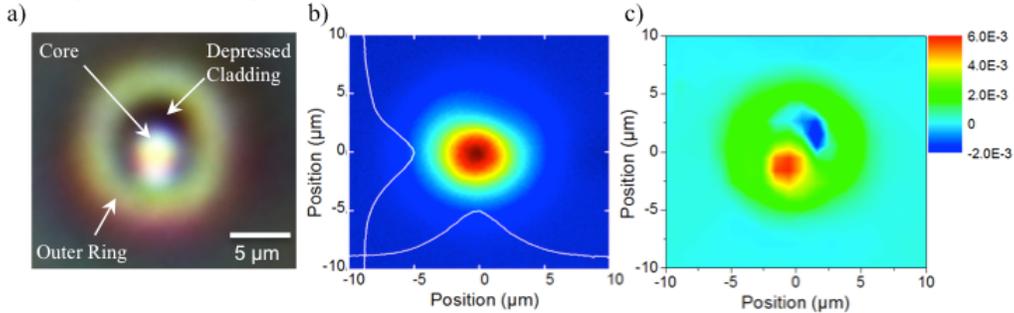

Fig. 1: (a) Micrograph image, (b) Mode field profile at 1550 nm and (c) Refractive index profile (at 635 nm) for a waveguide written with high repetition rates in Eagle2000 glass (pulse energy = 40 nJ).

To address the bend-loss limitations of high-repetition rate waveguides we have augmented the fabrication process. The technique now consists of a two-step process: the fabrication of large multimode waveguides (3-4 times larger in diameter than the usual single-mode waveguide) and a subsequent annealing stage. In the second step of the process, we exploit the different thermal stabilities between the various regions of the cumulative heating written structure. These different thermal stabilities enable the erasure of the structure's outer ring and the removal of residual laser induced stress. As a result, we obtain a single modified region of slightly smaller dimensions than that of a typical 40 nJ single-mode guide with two critical improvements: firstly, as the initial multimode guide is written with higher pulse energies, the remaining core has a larger index contrast (therefore, a smaller MFD can be achieved to further reduce the bend losses) and secondly, as the outer ring of index change is removed, the index profile post annealing becomes Gaussian-like in shape.

In 2008, *Eaton et al.* [15] reported the effect of thermal annealing on the physical dimensions and MFDs of waveguides written in the cumulative heating regime. However, to the best of our knowledge, there has not been any demonstration of improving waveguide performance by exploiting such techniques (or the different thermal stability regions of the glass). We demonstrate that by applying the process described above, waveguides with low bend losses at small radii become possible, bringing rapid prototyping of efficient 3D photonic devices closer to reality.

The paper is structured in the following way. In section 2 we will describe the fabrication technique and subsequent annealing procedure in detail. Section 3 summarizes the characterization of the waveguides and leads into a discussion while Section 4 shows a real world application. Concluding remarks are made in Section 5.

## 2. Fabrication

The waveguides were inscribed using an ultrafast Ti:sapphire oscillator (800 nm center wavelength) which emits laser pulses with a duration <50 fs at a repetition rate of 5.1 MHz (FEMTOSOURCE XL 500, Femtolasers GmbH). The circularly polarized laser beam was focused by a 100× oil immersion objective lens (Zeiss N-Achroplan, 1.25 Numerical Aperture, 450 µm physical working distance) into an alkaline earth boro-aluminosilicate glass substrate (Corning Eagle2000). The sample was placed on a set of 3-axis Aerotech air-bearing translation stages to achieve smooth translation during inscription. To measure the bend losses, arcs spanning 90° with radii ranging from 6.6 - 40 mm in eleven steps were inscribed into a sample with the dimensions 40×40×1.1 mm as seen in Fig. 2(a). Furthermore, the sample also contained 2 straight waveguides for propagation loss normalization purposes. Figure 1 shows the layout of the chip.

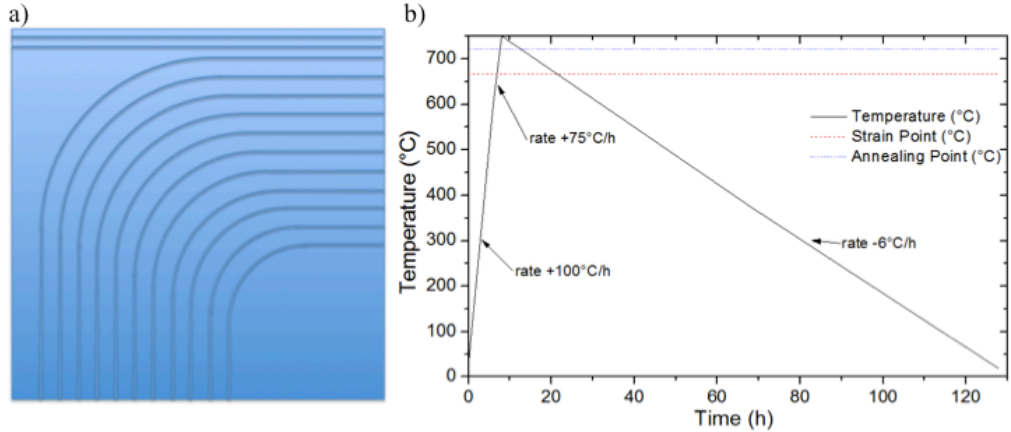

Fig. 2: (a) Schematic of the chip layouts. (b) Temperature profile for the rate annealing process. The process was done in 3 steps: first, heating the chip to 600 °C in 6 hours (rate: 100°C/h), then to 750°C in 2 additional hours (75°C/h) and last cooling the chip down to 18°C in 120 hours (-6 °C/h).

In order to fabricate single-mode waveguides at 1550 nm, the typical fabrication parameters are ~40 nJ pulse energy and 250-1500 mm/min translation velocities, resulting in circularly cross-sectioned waveguides with physical diameters of 12 μm as shown in Fig. 1. However, for the novel fabrication process presented here, waveguides were written with the greater pulse energy of 90 nJ and a translation speed of 500 mm/min, which resulted in large diameter (~30 μm) multimode waveguides. After fabrication the sample was ground and polished to reveal the waveguide ends.

Once the sample was polished, we administered a thermal treatment in order to erase the outer ring of the refractive index modification. A rate annealing process [17] was chosen for this purpose (in an air environment and room pressure), which is based on initially heating the sample above the transformation temperature in order to initiate an erasure and stabilization process and is shown in Fig. 2(b). Once the maximum temperature has been reached (which is above the annealing point: 722°C for Eagle2000), a very slow cooling gradient is applied until the glass is cooled below the strain point or transformation temperature (666°C), in order to ensure that the glass cools adiabatically. The key feature of this annealing process type is the slow cooling rate as it allows for the stress and birefringence to be removed as well. The profile we used consisted of heating the chip to 600 °C in 6 hours (rate: 100°C/h), then up to 750°C in an additional 2 hours (75°C/h) before cooling it down to 18°C in 120 hours (-6°C/h). Figure 2(b) shows the corresponding temperature profile. It is worth mentioning that this annealing procedure was administered at temperatures well below the softening point of the glass which is 985°C for Eagle2000.

After the thermal annealing process, waveguides were imaged under a transmission microscope (Olympus IX81) to analyze the physical changes produced. Bright field images before and after annealing are shown in Fig. 3(a) and 3(b) respectively. It can clearly be seen that the surrounding ring of index modification has been erased, leaving behind a waveguide that consists of a positive refractive index core surrounded by a narrow depressed region (dark ring) only. In order to determine whether the waveguides were truly single-mode, the mode emanating from the guides was imaged on a camera while the injection was offset in an attempt to excite higher order modes. Light at 1550 nm was launched into the waveguides using a SMF-28 fiber and the mode fields were imaged with an IR camera (Spiricon SP1550M). By using this technique it was determined that indeed the waveguides were single-mode. Figure 3(c) shows an image of a mode from a representative waveguide with a horizontal and vertical MFD of 8.5 and 10 μm, respectively (note that the subfigure has a different horizontal and vertical scaling as compared for Fig. 3(a) and (b)). As can be seen all the energy is confined within the central core of the waveguide. These MFDs result in a mode-mismatch coupling loss of 0.25 dB to a SMF-28 fiber calculated by numerically evaluating the mode overlap integral.

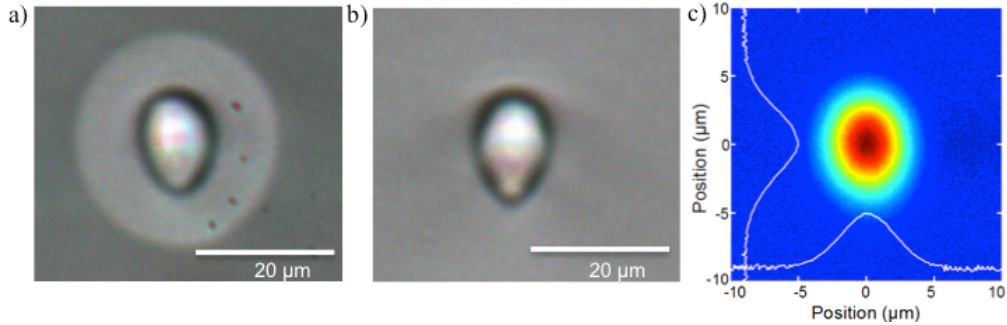

Fig. 3: (a) Bright field images of the waveguides before and (b) after annealing and the corresponding mode-field for the annealed waveguides (8.5 × 10 μm MFD) (c). Note that a different scale bar has been used in Figure 3(c) for better visualization.

3. Waveguide characterization

In order to quantify the physical changes induced by the writing laser and the influence of the differential post-annealing process, refractive index profiles were measured using a refracted near-field profilometer (Rinck Elektronik) at 635 nm before and after annealing (see Fig. 4). The initial refractive index profile exhibits a maximum index change of $9.7 \times 10^{-3}$ in the central region surrounded by a strip of $\Delta n = -1 \times 10^{-3}$, which is enclosed by a cladding with a nearly uniform index change of $\sim 3.2 \times 10^{-3}$. After annealing, it is clear that almost all traces of the outer cladding are gone (Fig. 4(b)). Furthermore the peak refractive index of the central region is reduced by $1.3 \times 10^{-3}$ to a peak value of $8.4 \times 10^{-3}$ along with the depressed region that has increased in magnitude to $\Delta n = -2 \times 10^{-3}$. The erasure of the outer refractive index modifications while the core index contrast is almost entirely preserved indicates that the structural changes of the glass network induced by the laser are of a different nature in the core and in the outer regions. This is related to the fact that the different regions are exposed to different thermal conditions during the fabrication process. *Little et al.* investigated the femtosecond laser induced structural changes in a multicomponent silicate glass (Schott BK7) in two different regimes [12]. Firstly, at low repetition rates, which corresponds to a repetitive modification of the glass network at lower temperatures and secondly at high repetition rates, which induces higher temperatures due to cumulative heating of the irradiated volume. In both regimes a positive refractive index change was found. However, at low repetition rates/low temperatures the origin of refractive index change was related to a change in polarizability, due to more ionic bonds, whereas at high repetition rates/high temperatures the origin of refractive index change was related to densification and rarefraction of the glass network [12]. This change of density is associated with the redistribution of elements, where network modifiers migrate out of the central region [18]. However, the temperatures involved in the post-annealing process are insufficient for reversing the elemental migrations [19] thereby preserving the core of the refractive index modification. Conversely the removal of the outer ring indicates that the associated index change is due to molar refractivity, which is susceptible to erasure at the temperatures reached in this annealing process [12].

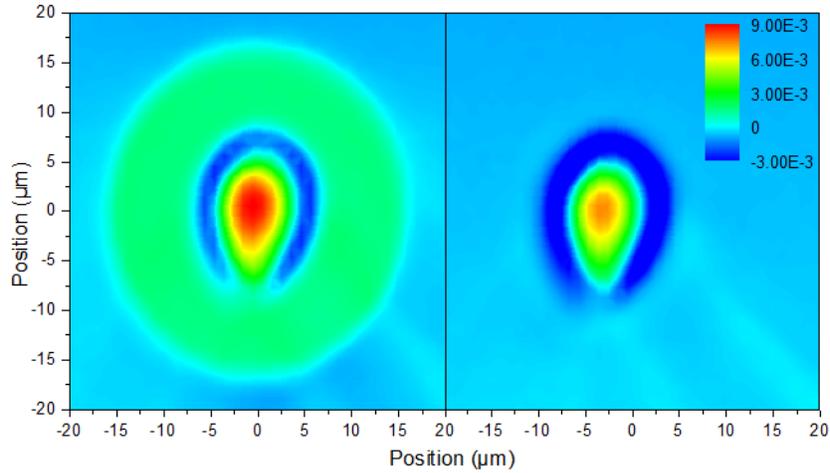

*Fig. 4: Refractive index profiles of a 90 nJ waveguide measured before (a) and after annealing (b).*

The throughputs of the chip were measured at 1550 nm by butt-coupling a single-mode fiber to the input of each waveguide using immersion oil while the light was collected with a power meter at the output end of the waveguide. A collection fiber was not used to route the light to the power meter at the output because the output plane of the waveguides was at 90° to the injection plane, which isolated the emerging signal from stray light in the block. The transmitted powers were normalized against the injected power and losses due to coupling, propagation and bulk glass absorption were subtracted in order to isolate the bend losses. The coupling loss between the waveguides and SMF-28 fiber was calculated to be $6 \pm 2\%$ by numerically evaluating the mode-overlap integral, and the absorption loss in Eagle2000 glass was accounted for with an absorption coefficient of $\alpha = 0.0075 \pm 0.0003$ mm$^{-1}$ at 1550 nm [20]. Finally, the bend losses were rescaled to a 30 mm length, which corresponds to the practical length of a device and are plotted in Fig. 5 (red dots). For comparison we have also replotted the data recently reported by *Charles et al.* [16], for a chip fabricated with identical specifications (number of guides, bend radii, length of guides, etc) which utilized non-annealed cumulative heating waveguides written at 40 nJ instead (black dots). The data from Charles et al. was normalized using the same procedure.

It can be seen that, for small bend radii (<20 mm), the throughput of the annealed 90 nJ waveguides is far superior to that of the non-annealed 40 nJ guides. In particular the 50% transmission point is found at a radius of 23.3 and 13.3 mm for the 40 and 90 nJ guides, respectively. We attribute this to two factors: Firstly, the central core region has a greater index contrast due to the higher pulse energies used for inscription. Secondly, the index profile now only consists of a Gaussian-like region, which has a similar performance to a step-index profile which is known to guide well around bends [21]. In addition for large bend radii (>20 mm), it can be seen that the throughput for the annealed waveguides plateaus at approximately 91%, which means that near-lossless soft bends can be used in circuit designs with minimal penalty. This was clearly not the case for the non-annealed guides which show greater losses even in this large bend radius regime.

To validate the trends in the data, bend loss simulations of the fabricated waveguides were conducted. A commercial beam-propagation software, BEAMProp by RSOFT, was used. For the simulations, the core radius of the waveguides was set to 4.85 µm, the wavelength to 1550 nm and the background refractive index of the glass to 1.4877 (the measured bulk refractive index of Eagle2000 at 1550 nm). The measured refractive index profiles of the annealed and non-annealed waveguides were then inserted into the software. The bend losses were modeled by using the "simulated bends" function available in the beam propagation tool. Results were then rescaled to a 30 mm length for comparison with the measured data. The simulated throughputs for the annealed and non-annealed guides are displayed in Fig. 5 as a solid red and black line respectively. For completeness we simulated a step-index waveguide/fiber with the same core size and an index contrast of $5.3 \times 10^{-3}$ (the index contrast of SMF-28 fiber). It can be seen that modeled data fits the measurements reasonably well. However, due to the complex nature of the index profile of the guides and the difficulty with calculating bend losses with such tight bend radii, there are some minor deviations. Finally as mentioned earlier the step-index profile offers the best bend loss performance, which the non-annealed waveguide most closely approaches due to its near-Gaussian index profile core.

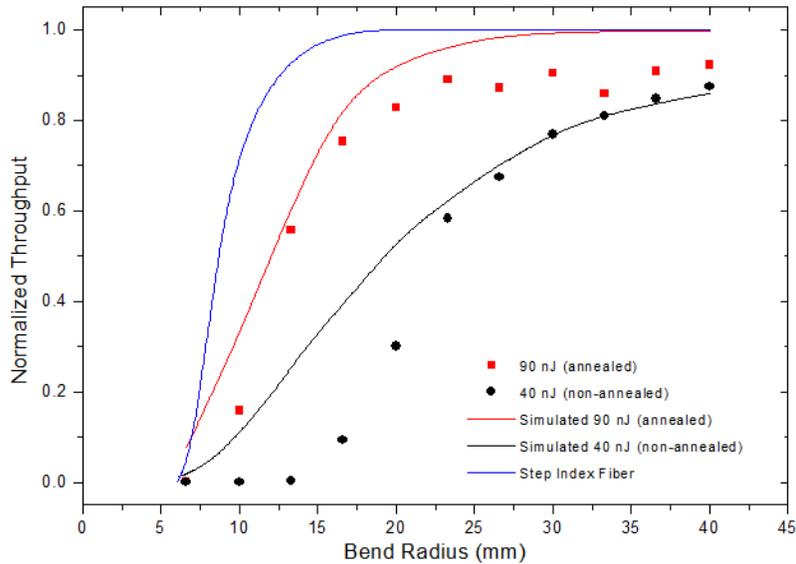

*Fig. 5: Normalized throughput as a function of radius of curvature for a 30 mm long waveguide. Solid lines show the result of BeamPROP simulations for the measured refractive index profiles. A step-index with a contrast of $5.3 \times 10^{-3}$ was used for the single mode fiber with a core radius of 4.85 µm.*

### 4. Real-world application: The integrated photonic pupil-remapper

To demonstrate the potential of these low-bend loss waveguides we applied them to the fabrication of an integrated device known as a "pupil-remapper" [10,16]. A pupil-remapper is a 3D photonic device that consists of a set of waveguides, which remap the light from a 2D input plane to a linear array at the output of the chip . The purpose of the pupil remapper is to reformat the light from the telescope pupil into a linear array so that all the light from the pupil can be interferometrically recombined downstream. The interferometric data is then used to reconstruct a diffraction-limited image of the stellar target despite of the presence of atmospheric turbulence. In this way it will be used to image protoplantary disks around stars in order to study large exoplanets during formation which will enhance our in our understanding of the formation and evolution process.

This astrophotonic device is subject to numerous constraints, which complicate its design; chief amongst these are maintaining a minimum distance between waveguides to avoid cross-coupling and equalizing path-lengths for high contrast fringe formation. The guides are also offset laterally from the injection position such that they do not overlap with the uncoupled cone angle of light from the injection micro-lenses. We refer the interested reader to [10] and [16] for full details about the design procedure and astronomical motivation. Recently, *Charles et al.* fabricated an 8 waveguide prototype pupil-remapper based on non-annealed cumulative heating waveguides [16], a diagram of which is depicted in Fig. 6(a). The guides in the remapper were based on circular arcs, which have constant bend radii along their lengths that ranged from 23-35 mm (See Fig. 6(c)). The prototype remapper was fabricated using the same writing-laser as described above with 40 nJ pulse energies and 250 mm/minute translation speeds without annealing. The device was 30 mm long and ~6.5 mm wide. The throughputs for the 8 waveguides of the device reported by *Charles et al.* were between 5 and 47%. Although these values are good enough for preliminary astronomical experiments, the dramatically unequal throughputs between pairs of waveguides used to form a given interferometric fringe result in a degradation of fringe contrast. Such loss in signal fidelity has a direct negative impact on the utility of such a device as a core component of a stellar interferometer.

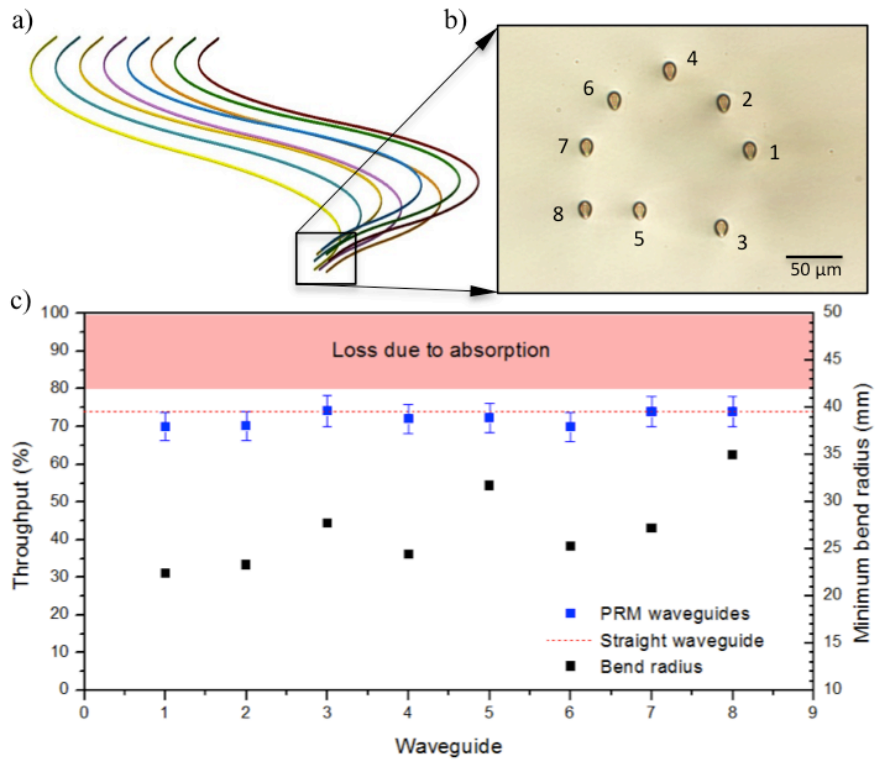

Fig. 6(a) Diagram of a 8 waveguide pupil-remapper (taken from [16]). The waveguides are spaced by 250 μm at the output of the chip. (b) Bright field microscope image of the input facet of the annealed remapper fabricated in this body of work. (c) Measured throughput and minimum bend radius of the individual waveguides within the pupil-remapper chip.

We fabricated the same device shown in Fig. 6, using the new fabrication method, namely by inscribing 90 nJ waveguides and then subsequently applying the annealing process outlined above. The sample was ground and polished prior to probing. A microscope image of the input facet of the fabricated device is shown in Fig. 6(b). The 1550 nm laser probe was injected into each guide with a SMF-28 fiber and collected at the output by another fiber before being routed to a power meter. The throughput of a straight reference guide, embedded in the block adjacent to the remapper, was also measured. All throughputs were normalized to two butt-coupled probe fibers without the pupil-remapping chip in between. The absolute throughputs of the 8 waveguides of the remapper were between 70 and 73% (see Fig. 6c), which constitutes a major improvement over the device reported by *Charles et al.* The remaining losses are mainly attributed to bulk absorption in the 30 mm long Eagle2000 block that contributed 20% (highlighted by a red box in Fig. 6) to the loss and two 6% losses (per facet) due to the mode-mismatch between the fibers and the waveguide mode at each end of the device. Indeed, when the 8 remapper throughputs were normalized with respect to the straight guides (dashed red line), the throughputs ranged between 96 and 100%. This indicates that there were negligible bend losses as is expected for 30 mm long waveguides with bend radii between 23 and 35 mm from Fig. 5. As the bend losses have now been minimized, it now becomes possible to redesign a remapper which has a smaller footprint (which will have shorter tracks and hence lower absorption losses), by increasing the minimum bend radius from the 23 mm used in the device discussed here towards 17 mm which sits at the edge of the throughput curve in Fig. 5. There is a limit to this as well and we estimate that a gain of 5-8% improvement in throughput can be made this way.

The vastly improved and equalized throughputs of the remapper will significantly increase the fringe contrast of an interferometer based on this chip and allow for very precise determination of the complex coherence of the incident radiation field which is key to high fidelity astronomical imaging (with specific application to high contrast detection of faint exoplanets against the glare of their host stars). However, measurements of the precision of such an interferometer are deferred to future work. This illustrates a successful real-world application of this technique.

## 5. Conclusions

In conclusion, we have identified different thermal stability regions in the direct-written waveguides and have utilized a differential thermal annealing process to significantly decrease the bend losses of waveguides written in the cumulative heating regime. The thermal annealing process exploits the different thermal stability of different regions within the induced refractive index profile. The heat treatment erases the outer ring, but leaves a very high index contrast ($8.4\times10^{-3}$), Gaussian-like profile core behind. As a result, the waveguides show dramatically superior bend loss characteristics as compared to non-annealed waveguides. Indeed, we have demonstrated that this method has enabled the realization of an efficient 8 waveguide 3D, pupil-remapping chip. This performance improvement will have far reaching implications allowing rapid prototyping with MHz repetition rates (fabrication speeds in the range of 500-2000 mm/minute), while retaining the ability to fabricate complex, multi-element 3D photonic circuits.

**Acknowledgements**

This research was conducted by the Australian Research Council Centre of Excellence for Ultrahigh bandwidth Devices for Optical Systems (project number CE110001018) and in part performed at the Optofab node of the Australian National Fabrication Facility; a company established under the National Collaborative Research Infrastructure Strategy to provide nano and microfabrication facilities for Australia's researchers.